\documentclass[aps,prx,twocolumn,superscriptaddress]{revtex4-2}

\usepackage{mathrsfs}
\usepackage{amsmath}
\usepackage{amssymb}
\usepackage{bm}
\usepackage{graphicx}
\usepackage{hyperref}
\usepackage{physics}
\usepackage{braket}
\usepackage{xcolor}
\usepackage{siunitx}
\usepackage[caption=false]{subfig}
\usepackage{dsfont}

\definecolor{darkgreen}{rgb}{0.0, 0.5, 0.0}

\hypersetup{
    colorlinks=true,
    linkcolor=blue,
    citecolor=blue,
    urlcolor=blue
}

\begin{document}

\title{Emergent spatiotemporal order and nonreciprocity in \\ driven-dissipative nonlinear magnetic systems}
\author{Vincent P. Flynn}
\affiliation{Department of Physics, Boston College, Chestnut Hill, Massachusetts 02467, USA}

\affiliation{Department of Physics and Astronomy, Dartmouth College, Hanover, New Hampshire 03755, USA}

\author{Benedetta Flebus}
\affiliation{Department of Physics, Boston College, Chestnut Hill, Massachusetts 02467, USA}

\date{\today}

\begin{abstract}
The identification of platforms with independently tunable nonlinearity and non-Hermiticity promises a quantitative route to far-from-equilibrium universality across many-body systems. Here we show that a conventional ferromagnetic multilayer realizes this paradigm: balancing a dc drive against Gilbert damping stabilizes a self-organized, current-carrying nonequilibrium condensate that spontaneously breaks spacetime-translation symmetry. The chirality of this spin superfluid limit cycle state generates an inherently nonreciprocal flow: long-wavelength magnons of opposite chirality acquire asymmetric dispersions and propagate direction-selectively, realizing a spin superfluid diode. This asymmetry is flow-borne -- it reflects broken Galilean invariance and requires neither structural asymmetry nor finely tuned gain–loss balance. Linearized dynamics in the comoving superfluid frame are intrinsically pseudo-Hermitian and, in the long-wavelength sector, can be mapped to a (1+1)D wave equation on curved spacetime. Spatial modulation of the drive enables the generation of sonic horizons that parametrically amplify magnon pairs and produce Hawking-like particle–hole emission. Our results establish a tabletop route from nonlinear dissipative-driven magnetization dynamics to nonreciprocal transport,  nonequilibrium phase transitions, and analogue-gravity kinematics.
\end{abstract}

\maketitle

\section{Introduction}

Most of the surrounding world operates far from equilibrium. In this regime, the interplay between nonlinear and energy non-conserving interactions is an organizing principle that cuts across seemingly unrelated phenomena -- from  neural networks \cite{Neural_Chialvo} to
astrophysical flows \cite{Percolation_Schulman} -- where collective order is set by symmetries and conservation laws rather than microscopic detail, enabling sharp tests of universality.

Dissipative Rydberg arrays and exciton–polariton fluids have provided early platforms for quantitative tests that have established a paradigm for exploring universal behavior in  open quantum systems \cite{RybergCount_Malossi2014,SelfOsccQED_Dreon2022,KPZPolariton_Fontaine2022,Universality_Diehl2025}.  The  pace of theoretical advances, however, calls for platforms in which nonlinearity and non-Hermiticity can be easily tuned and precisely diagnosed with tabletop instrumentation.  Such systems could enable parameter-resolved maps of dynamical phase diagrams and tests of universality, while opening access to properties unique to far-from-equilibrium physics such as critical behavior organized not by equilibrium free-energy landscapes, but by spectral singularities forbidden at thermal equilibrium \cite{DPT_Diehl2010,DPT_Mattias2017,SpectralDPT_Minganti2018,NHPT_Hanai2019,CritEP_Hanai2020,NonRec_Vitelli2021,NLStochEP_Vitelli2022,QBLCrit_Yikang2022,IrrevFluc_Suchanek2023,Bosoranas_2023,MariamPRA_2024,CEP_Diehl2024,NRIsing_Vitelli2025}. 
 
With their intrinsic dissipative and nonlinear dynamics~\cite{deng2023exceptional}, and well-established protocols for spin injection, magnetic systems appear as promising, low-overhead solid-state candidates. Yet most explorations of their non-Hermitian properties have remained confined to the linear regime \cite{MagLossGain_Lee2015,NHTopMag_McClarty2019,TopoMagAmp_Nunnenkamp2019,NHSEMag_BF2022,NHmag_BF2022,MagZM_Gunnink2023,RecentMagnon_BF2023,UnidirectMagnon_Rembert2023,OscEdge_Rembert2024,DissSWDiode_Zou2024}.

\begin{figure}[h!]
    \centering
    \includegraphics[width=\linewidth]{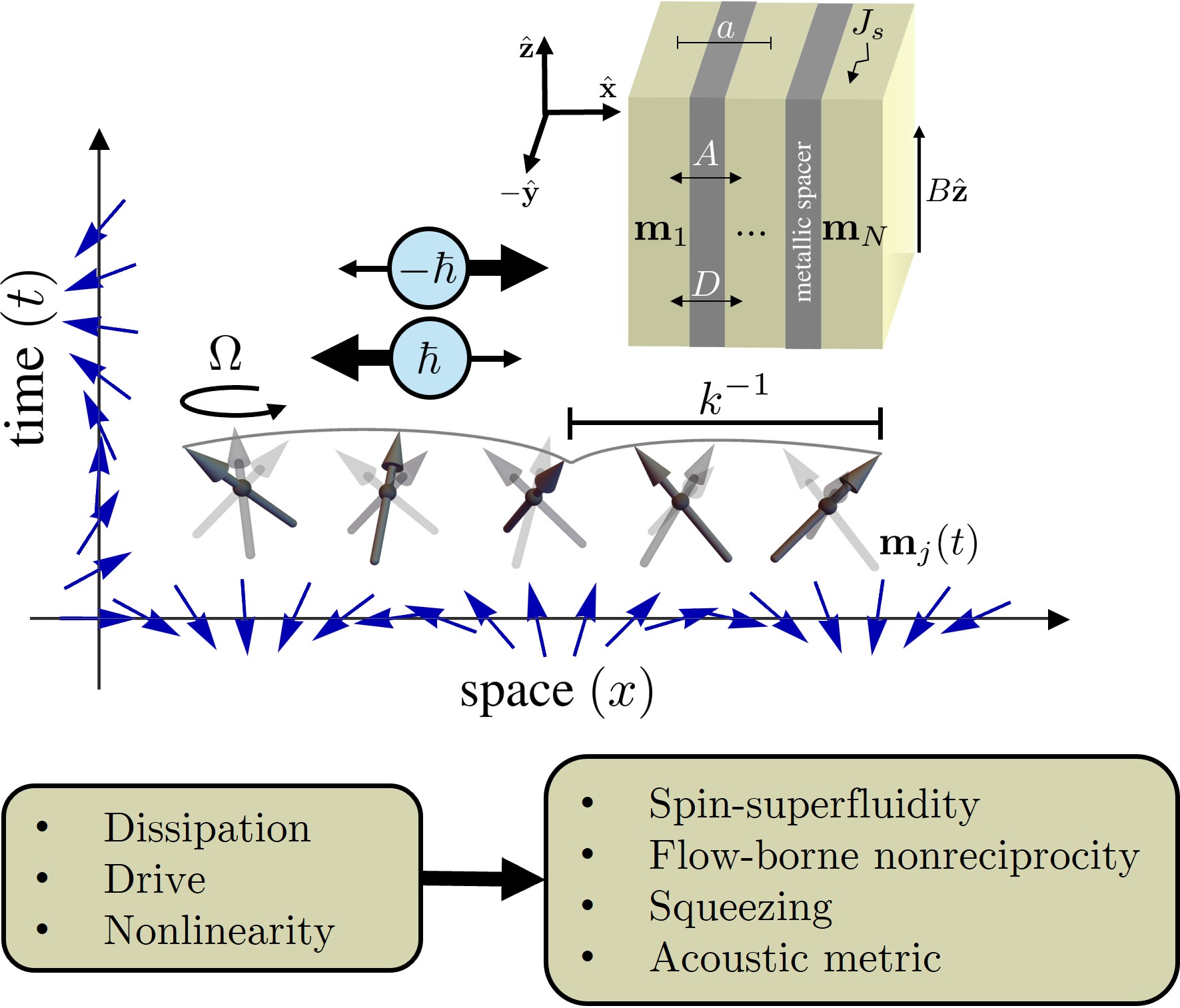}
    \caption{A schematic depiction of the superfluid limit cycle (SFLC) state realized in our model and the resulting spin wave dynamics. The state breaks both spatial and temporal translation symmetry and supports excitations of opposite angular momenta with asymmetric propagation speeds. Inset: a schematic of the heterostructure device wherein the SFLC can be realized. Each ferromagnetic layer is subjected to the applied field $B\hat{\mathbf{z}}$ and separated by a metallic spacer that mediates both interfacial symmetric and antisymmetric (DM) exchange $A$ and $D$, respectively. An antidamping torque of strength $J_s$ counteracts the intrinsic dissipation of the magnetization dynamics.   }
    \label{fig: SFLC_cartoon}
\end{figure}

Here we move beyond this regime and demonstrate that the interplay of drive, dissipation, and coherent interactions in a conventional ferromagnetic heterostructure can dynamically stabilize a steady state with no equilibrium analogue: a self-organized, current-carrying nonequilibrium condensate. Specifically, we show that balancing Gilbert damping with a dc drive in the magnetic multilayer depicted in Fig. \ref{fig: SFLC_cartoon}  allows for the emergence of a spacetime-translation symmetry-breaking chiral spin superfluid limit cycle. The resulting superflow provides an emergent nonreciprocal background: long-wavelength magnons of opposite chirality acquire intrinsically asymmetric dispersions and propagate direction-selectively, realizing a spin superfluid diode. Crucially, the nonreciprocity is flow-borne: it is neither inherited from built-in structural asymmetries nor reliant on fine-tuned balancing of coherent and dissipative interactions -- it originates from the broken Galilean invariance.

We further establish a concrete analogue-gravity link by showing that, in the comoving, corotating frame of the spin superfluid limit cycle, phase fluctuations satisfy a scalar wave equation with an effective (1+1)D acoustic metric. The chiral background 
acts as a parametric pump for magnon pairs - the classical parametric amplification mechanism that, upon quantization, produces a stationary two-mode squeezed state. Introducing a current-modulation junction nucleates a sonic horizon across which one mode tunnels by particle–hole pair production while its negative-energy partner is advected, yielding Hawking-like emission bursts.

Taken together, our results elevate a seemingly mundane magnetic heterostructure to a platform in which spin superfluidity, emergent nonreciprocity and Hawking-like emission can be generated, tuned, and diagnosed with standard instrumentation.

\section{Model}
Our starting point is a heterostructure comprising $N$ ferromagnetic layers separated by metallic spacers of thickness $a$, which mediate an effective  RKKY-like exchange stiffness $A>0$ and Dzyaloshinskii–Moriya interaction (DMI) $D\in\mathbb{R}$ between neighboring layer magnetizations $\mathbf{m}_j$ and $ \mathbf{m}_{j\pm 1}$. An external magnetic field $\mathbf{B} = B \hat{\mathbf{z}}$ aligns the static equilibrium magnetization along the $\hat{\mathbf{z}}$ axis. Each layer obeys Landau–Lifshitz–Gilbert  (LLG) dynamics characterized by an intrinsic Gilbert damping \(\alpha\) and an antidamping Slonczewski torque generated via a dc spin current $\mathbf{J}_s=J_s \hat{\mathbf{z}}$ \footnote{
We retain only the damping-like (Slonczewski) torque. A field-like component $\propto \mathbf{m}\times \hat{\mathbf{z}}$ collinear with the applied field simply renormalizes the precession frequency and does not qualitatively affect the conclusions of this work.
}. Heterostructures hosting such interactions have been both explored  theoretically and realized experimentally~\cite{STT_Slon1996,GilbLayer_Berger2001,EnhancedGilb_Yaro2002,Pumping_Yaro2002,DMIObs_Di2015,LongRangeChiralAFM_Han2019,LargeDMI_Arregi2023,XinPaper_2025}. 

To make our analysis physically transparent, we adopt a continuum description, coarse-graining $\mathbf{m}_j(t) = \mathbf{m}(x,t)$, with $x=ja$ - see Appendix \ref{app: num} for details on the discrete structure.  In the continuum limit, the state of the system is described by a spacetime-local unit vector $\mathbf{m}(x,t)$ evolving under the LLG equation
\begin{align}\label{eq: LLG}
    \partial_t\mathbf{m} = -\mathbf{m}\times\mathbf{h} + \alpha\mathbf{m}\times\partial_t\mathbf{m} + J_s\mathbf{m}\times(\mathbf{m}\times\hat{\mathbf{z}})\,,
\end{align}
where time is measured in units of the characteristic precession frequency $\omega=\gamma B$, with $\gamma$ the gyromagnetic ratio. The dimensionless effective field $\mathbf{h} = -\delta \tilde{F}/\delta \mathbf{m}$ is derived from the continuum free energy functional \(F[\mathbf{m}]=BM_s\tilde{F}[\mathbf{m}]\), with $M_s$ the saturation magnetization and
\begin{align}\label{eq: Fdiml}
\tilde{F} = \int \left( \frac{1}{2} |D_x\mathbf{m}|^2 + \frac{d^2}{2}(m^z{}^2-1) -  m^z\right) dx,
\end{align}
the dimensionless free energy. Here,
$D_x\mathbf{m} = \partial_x\mathbf{m} +d\hat{\mathbf{z}}\times\mathbf{m}$ is a covariant derivative associated with an effective $\text{U}(1)$ gauge field of strength $d=\ell_B/\ell_D=D/(AM_s B)^{1/2}$, where $\ell_B = (A/M_sB)^{1/2}$ and $\ell_D=A/D$ are characteristics lengths associated with the 
exchange strength $A$ and DMI strength $D$, respectively. We have also rescaled $x\mapsto \ell_B x$. This covariant formulation of the DMI reveals that, in a frame cospiralling with the DMI-imposed twist (wherein $D_x\mapsto \partial_x$), the  magnetization experiences an effective anisotropy of strength $d^2/2$  favoring $\mathbf{m}\perp \hat{\mathbf{z}}$ \cite{GaugeSpin_Tatara2019,ChiralGeom_Hill2021,GaugeMag_DiPietro2022,GaugeChiral_Ansalone2022}.

While the model has a global U(1) symmetry about $\hat{\mathbf{z}}$, it lacks a local one: under a position-dependent rotation by an angle $\varphi(x)$ about $\hat{\mathbf{z}}$,  the DMI acts as a background gauge potential that transforms as $d\mapsto d+\partial_x\varphi$, thus exhibiting the minimal-coupling structure familiar from superfluid helimagnets~\cite{SFReview_Sonin,QHbilayer_Girvin}. 
Writing $\mathbf{m} = (\sqrt{1-m^z{}^2}\cos\phi,\sqrt{1-m^z{}^2}\sin\phi,m^z)$, global U(1) symmetry  yields the conserved Noether current
\begin{align}\label{eq: supercurrent}
    j(x) = D_x\mathbf{m}\cdot(\hat{\mathbf{z}}\times\mathbf{m}) = (1-m^z{}^2)(\partial_x \phi + d)\,.
\end{align}
Equation \eqref{eq: supercurrent} defines the spin superfluid current carried by the azimuthal phase $\phi$. In equilibrium, the free energy \eqref{eq: Fdiml} admits stationary states with uniform current \(j(x)=j_0\) corresponding to spiral states (SSs) of pitch \(\partial_x\phi=k\) and $m^z=-1/f_k$, with $f_k = k(k+2d)$ the effective stiffness associated with a spiral of pitch $k$. These SSs are thus entirely specified by $k$.

\begin{figure}
    \centering
\includegraphics[width=\linewidth]{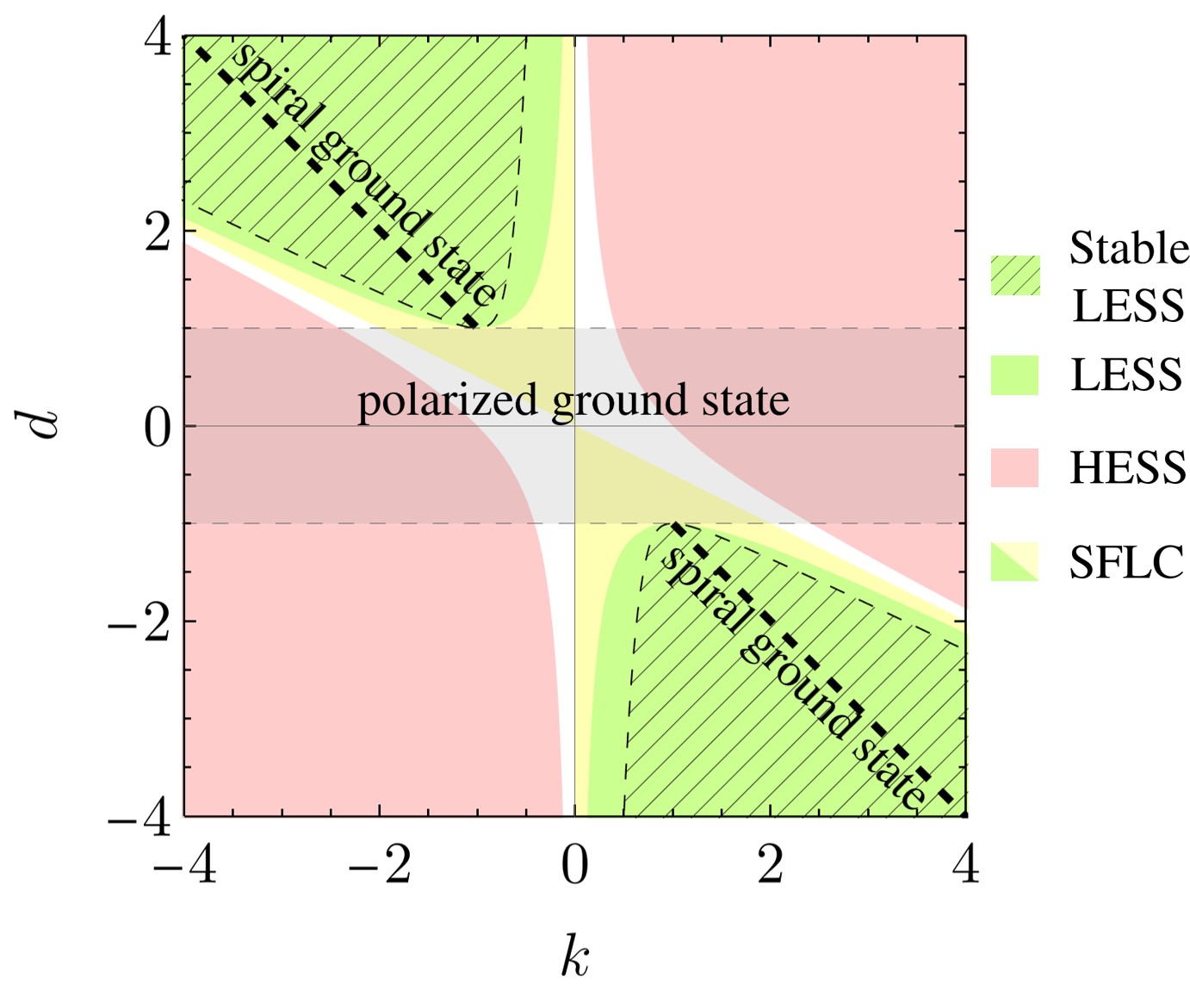}
    \caption{ The equilibrium phase diagram of the model in Eq.\,\eqref{eq: Fdiml} superimposed with that of the SFLC. Given a dimensionless DMI $d$ and spiral pitch $k$, the color of the corresponding region indicates the existence (or non-existence) of either an energetically stable LESS (hatched green), an unstable LESS (unhatched green), or a HESS (red) - which are always unstable (see \ref{app: eqanalysis} for more details). The grayed out region corresponds to $|d|<1$ whereby the ground state is fully polarized $\mathbf{m}=\hat{\mathbf{z}}$. Outside of this region, the ground state is the stable LESS with pitch $k=-d$ indicated with the black dashed line. The yellow/green region corresponds to those values of $d$ (and corresponding values of $k$) where an SFLC of pitch $k$ can be stably supported. Crucially, the yellow region extends to the weak-DMI regime $|d|<1$ where equilibrium superfluid states are forbidden.  }
    \label{fig: EqPD}
\end{figure}

Sufficiently large DMI renders some spiral states energetically stable, as shown by the equilibrium phase diagram in Fig.\,\ref{fig: EqPD}. Specifically,  we find that the stable uniform-current states with $k\neq -d$ (and thus $j_0\neq 0$) emerge as low-energy spiral states (LESSs) whenever $|d|>1$ with the LESS of pitch $k=-d$ (and thus $j_0=0)$ serving as the ground state (see \ref{app: eqanalysis}). 
Yet -- echoing the  proposals for spin superfluidity in easy-plane collinear magnets \cite{SF_Halperin1969,SF_Sonin1982,SFSpintron_Sun2016,SFReview_Sonin} -- an unambiguous equilibrium realization remains experimentally elusive, suggesting that this current-carrying state is difficult to stabilize as a true ground state. We therefore adopt a dynamical approach: by counteracting damping with a dc drive, we seek to stabilize the same current-carrying configuration as a 
limit cycle  in the regime $|d|<1$, 
placing the emergence of spin superfluidity within experimental reach.

To make this concrete, we search for  steady, current-carrying solutions of Eq.\,\eqref{eq: LLG}. Introducing the canonically conjugate variables ($m^z$, $\phi$) allows us to recast the dynamics in hydrodynamic form, yielding the coupled nonlinear equations

\begin{subequations}\label{eq: mzphieom}
  \begin{align}
   \partial_t m^z + \partial_x j &= (1-m^z{}^2)(\alpha \partial_t\phi - J_s)\,, \label{cont}
   \\
   \partial_t\phi &= -\frac{\delta \tilde{F}}{\delta m^z}-\alpha\frac{\partial_t m^z}{1-m^z{}^2}\,.\label{phase}
\end{align}  
\end{subequations}
Equation~\eqref{cont} is naturally interpreted as a local spin continuity law: the divergence of the spin current balances nonconservative torques, with $J_s$ and $\alpha \partial_t\phi$, respectively, injecting and dissipating spin angular momentum. 
For a spatially uniform longitudinal component, i.e., $\partial_x m^z = 0$, Eq.~\eqref{phase} reduces to

\begin{align}
       \partial_t\phi &= 1+m^z f_\phi-\alpha\frac{\partial_t m^z}{1-m^z{}^2}\,,
\end{align}
where $f_\phi = (\partial_x\phi)(\partial_x\phi+2d)$ is the phase-stiffness density associated with a twist in  $\phi$. Equations~\eqref{cont} and~\eqref{phase} admit a uniformly precessing steady-state \textit{limit cycle (LC) solution} defined by
\begin{align}
    \phi_\text{LC}(x,t) &= kx + \Omega t\,, \quad 
    m^z_\text{LC} = \frac{\Omega - 1}{f_k}\,, \label{LCsol}
\end{align}
where $\Omega = J_s/\alpha$ is the precession frequency. In this state, the linear phase winding generates  a steady, nonzero flow of spin angular momentum, i.e., the spin supercurrent $j_k = n_k v_k$, with $n_k=1 - m^z_\text{LC}{}^2$ and $v_k=k+d$ defining the spin superfluid density and velocity, respectively. This nonequilibrium spin superfluid LC (SFLC) phase is well-defined provided that $|m^z_\text{LC}| < 1$, a condition generically met by a suitable choice of  $\Omega$ for any $k \neq 0,-2d$. Importantly, no restriction (other than $d\neq 0$) is placed on the DMI strength $d$, thus extending the regime of realizable stable superfluid states beyond what is possible in equilibrium (see Fig.\,\ref{fig: EqPD}). Importantly, discrete realizations of these solutions exist in realistic heterostructures - see Appendix \ref{app: num}.

Symmetry-wise, the SFLC solution~\eqref{LCsol} is highly nontrivial:  it dynamically breaks rotational, spatial, and continuous time-translation symmetries, exhibiting hallmarks of spacetime-crystalline order intertwined with superfluid transport \cite{TC_Wilczek2012,TCRev_Sacha2017,DissTCRydberg_Wu}. In particular, continuous space- and time-translational symmetries are broken down to discrete ones with periodicities $2\pi/k$ and $2\pi/\Omega$, respectively. The three broken symmetries, however, are not independent: the original 3-parameter symmetry group  is broken down to a single 2-parameter symmetry transformation $(x,t,\phi)\mapsto (x+a,t+\tau,\phi-ka-\Omega\tau)$. Consequentially, only $3-2=1$ Goldstone mode emerges associated with azimuthal phase fluctuations $\phi\mapsto \phi+\delta \phi$. The SFLC thus realizes a nonequilibrium incarnation of the so-called ``inverse Higgs mechanism", whereby the number of Goldstone modes is not equal to the number of broken symmetries \cite{InverseHiggs_Watanabe2014}.

The state is chiral: the phase propagates unidirectionally with velocity $-\Omega/d$, whose sign is set by $d$. Its existence, however, does not guarantee its dynamical stability: trajectories nearby need not converge to the solution. To identify the stable sector and the fluctuation spectrum, we proceed to linearize the dynamics about
 this chiral SFLC background. 
 
\section{Stability and excitations \\ of the SFLC state}
\begin{figure}
        \centering
        \includegraphics[width=.8\linewidth]{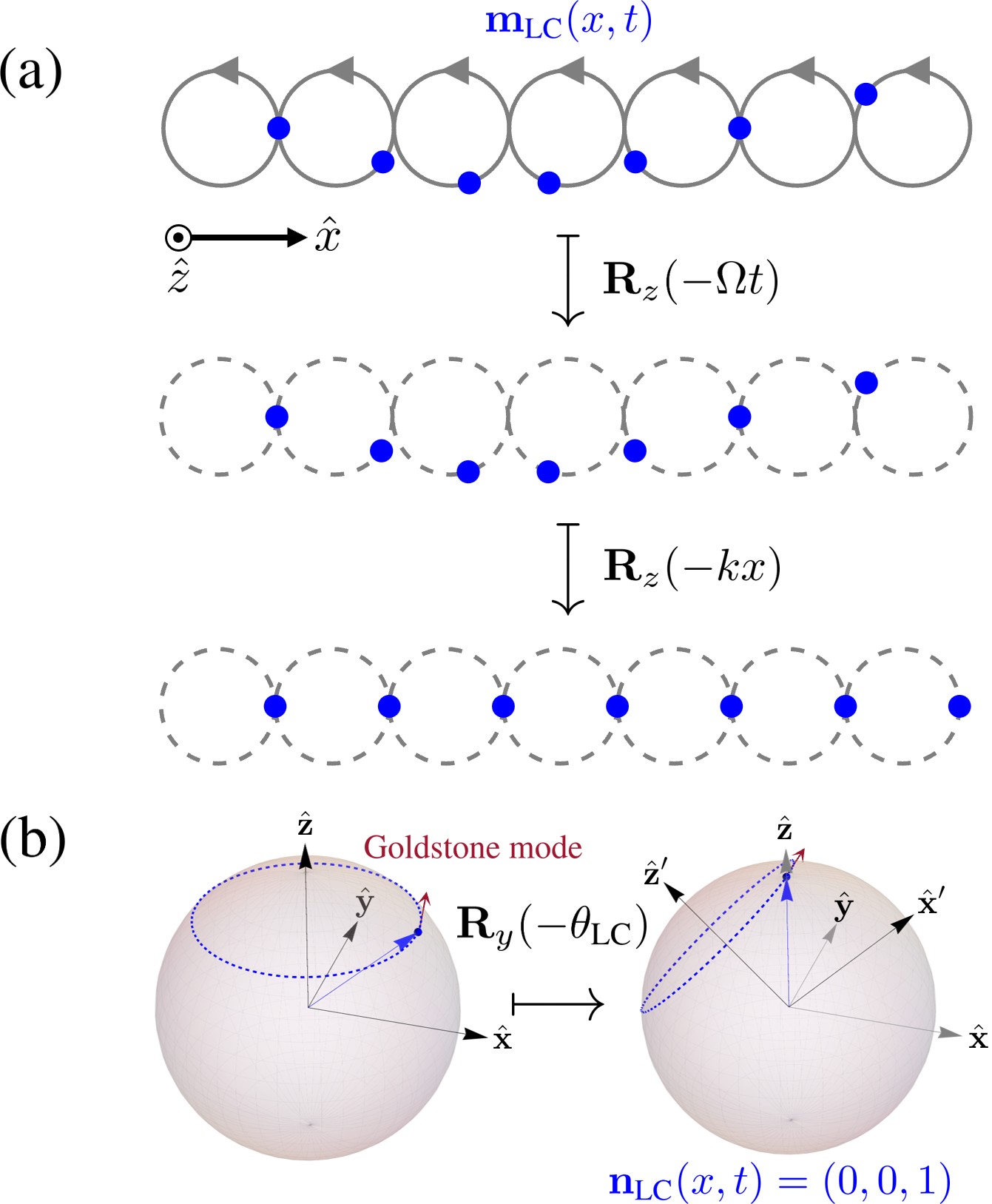}
        \caption{The SFLC unwinding map defined by Eq.\,\eqref{eq: unwind}. \textbf{(a)} The transverse components of $\mathbf{m}_{\text{LC}}(x,t)$ are shown in blue. The relative spatial variations are set by the SFLC pitch $k$. The gray arrow indicates the precessing dynamics at frequency $\Omega$ of the SFLC state. The first step of the unwinding map moves the system into a frame rotating at $\Omega$, rendering each spin stationary. The second step unwinds the spatial spiral of pitch $k$. \textbf{(b)} The magnetization is finally globally rotated so that the canted state (with polar angle $\cos\theta_\text{LC} = m_\text{LC}^z$)  is aligned along the north pole. The Goldstone mode associated with phase fluctuations $\phi\mapsto \phi+\delta\phi$ maps onto transverse fluctuations $n^y\mapsto n^y+\delta\phi\,\sin\theta_\text{LC}$, indicated by a red arrow.}
        \label{fig: UnwindFig}
    \end{figure}

In contrast to the usual equilibrium expansion, our background is an explicitly nonequilibrium limit cycle that breaks rotational, spatial, and continuous time-translation symmetry. To handle its complexity, we introduce a unwinding map that flattens the SFLC into a time-independent uniform polarized state amenable to conventional spin wave analysis. Specifically, we perform a spacetime-local SO(3) transformation $\mathbf{n}(x,t) = \mathbf{R}(x,t)\mathbf{m}(x,t)$ with
\begin{align}\label{eq: unwind}
    \mathbf{R}(x,t)&=\mathbf{R}_y(-\theta_\text{LC})\mathbf{R}_z(-kx)\mathbf{R}_z(-\Omega t)\,, 
\end{align}
where $\mathbf{R}_\mu(\theta)$ is the rotation matrix by an angle $\theta$ about the $\mu$'th Cartesian axis and $\cos\theta_\text{LC} = m_\text{LC}^z$. If $\mathbf{m}_\text{LC}(x,t)$ is the SFLC in vector form, then, by construction, $\mathbf{n}_\text{LC}(x,t)= \mathbf{R}(x,t) \mathbf{m}_\text{LC}(x,t) = \hat{\mathbf{z}}$. 
Figure \ref{fig: UnwindFig} illustrates the unwinding map defined by Eq.~\eqref{eq: unwind}. First, the dynamical rotation $\mathbf{R}_z(-\Omega t)$ freezes the uniform precession; next, the spatial unwinding $\mathbf{R}_z(-kx)$ unwinds the spatial helix; finally, the global tilt $\mathbf{R}_y(-\theta_{\mathrm{LC}})$ aligns the magnetization with the north pole.
In the comoving, corotating frame, the LLG equation \eqref{eq: LLG} and the free energy \eqref{eq: Fdiml}   can be recast in terms of the transformed magnetization $\mathbf{n}(x,t)$ via the substitutions:  
\begin{align}
    \hat{\mathbf{z}} &\mapsto \hat{\mathbf{z}}',\quad 
    \partial_t \mapsto \partial_t + \Omega \hat{\mathbf{z}}' \times, \quad
    D_x \mapsto \partial_x + (k+d) \hat{\mathbf{z}}' \times\,, \label{transf}
\end{align}  
where $\hat{\mathbf{z}}' = \mathbf{R}_y(-\theta_\text{LC})\hat{\mathbf{z}}$ (shown in Fig.\,\ref{fig: UnwindFig}(b)).
The first replacement fixes the tilted quantization axis set by the global $y$–rotation; the latter two encode the corotating/comoving $\mathfrak{so}(3)$ gauge potential generated by $\mathbf{R}(x,t)$, i.e., $(A_t,A_x)=(\Omega\,\hat{\mathbf z}'\!\times,\,k\,\hat{\mathbf z}'\!\times)$. The transformation~\eqref{transf} leads to the rescaling  
\begin{align}\label{eq: unwindrescale}
    d \mapsto d+k,\quad J_s\mapsto J_s-\alpha\Omega = 0\,.
\end{align}
Interestingly, Eq.\,\eqref{eq: unwindrescale} identifies $k=-d$ (corresponding to the LESS ground state in the strong DMI regime) as a gauge-neutral point: in the comoving frame the spatial $\mathfrak{so}(3)$ gauge field $A_x\propto (k+d)\,\hat{\mathbf z}'\!\times$ vanishes, the DMI shift cancels $(d\!\to\! d+k=0)$, and the superflow velocity is zero $(v_k=k+d=0)$, rendering the system equivalent to an achiral easy-plane ferromagnet. Importantly, under Eq.\,\eqref{eq: unwind}, the Goldstone mode associated with azimuthal phase fluctuations maps onto the transverse component $n^y$ which transforms as $n^y\mapsto n^y+\delta\phi\,\sin\theta_\text{LC}$ under $\phi\mapsto \phi+\delta\phi$ (see \ref{app: lab}).

Linearizing about the unwound background, we write $\mathbf n=\hat{\mathbf z}+\delta\mathbf n$ and define $\psi=\delta n^x + i\,\delta n^y$. In this frame, the original U(1) symmetry axis is shifted $\hat{\mathbf{z}}\mapsto \hat{\mathbf{z}}'$
(see Fig.\,\ref{fig: UnwindFig}(b)). Consequently, the quadratic theory about the new $\hat{\mathbf{z}}$ is not number-conserving and dynamically couples
$\psi$ with $\psi^\ast$. The fluctuation dynamics are therefore naturally cast in Bogoliubov--de Gennes (BdG) form:

\begin{align}\label{eq: dissBdG}
    i(\mathds{1}_2-i\alpha\bm{\sigma}_3)\partial_t\Psi_q = \mathbf{G}_q\Psi_q,
\end{align}
where $\Psi_q=(\psi_q,\psi^*_{-q})$ is the momentum-space Nambu spinor \footnote{This spinor characterizes magnonic excitations which, due to bosonic statistics, transform under SU(1,1), not SU(2), like conventional Nambu-Dirac spinors. These transformations preserve the norm $|u|^2-|v|^2$.} of fluctuations, $\bm{\sigma}_j$ are the Pauli matrices, and
\begin{align}\label{eq: Gq}
    \mathbf{G}_q = -2m_\text{LC}^z v_kq\mathds{1}_2 + (\Delta_k-q^2)\bm{\sigma}_3 + \Delta_k i\bm{\sigma}_2\,,
\end{align}
where  $\Delta_k = f_kn_k/2$ is the effective pairing strength, i.e., a parametric drive of particle–hole pairs. The effective BdG Hamiltonian $\mathbf{G}_q$ is \textit{pseudo-Hermitian} with respect to $\bm{\sigma}_3$, i.e., $\mathbf{G}_q^\dag = \bm{\sigma}_3 \mathbf{G}_q\bm{\sigma}_3$, reflecting the bosonic nature of magnonic excitations \cite{TopoMagnon_Mook2013}. Additionally, it satisfies the ``particle-hole" constraint $\mathbf{G}_q^* = -\bm{\sigma}_1\mathbf{G}_{-q}\bm{\sigma}_1$ \cite{Squaring}. Thus, in the formal limit where the dissipative term on the left-hand side of Eq.\,\eqref{eq: dissBdG} is set to zero, the nonequilibrium dynamics are effectively coherent.

\begin{figure*}
    \centering
    \includegraphics[width=\linewidth]{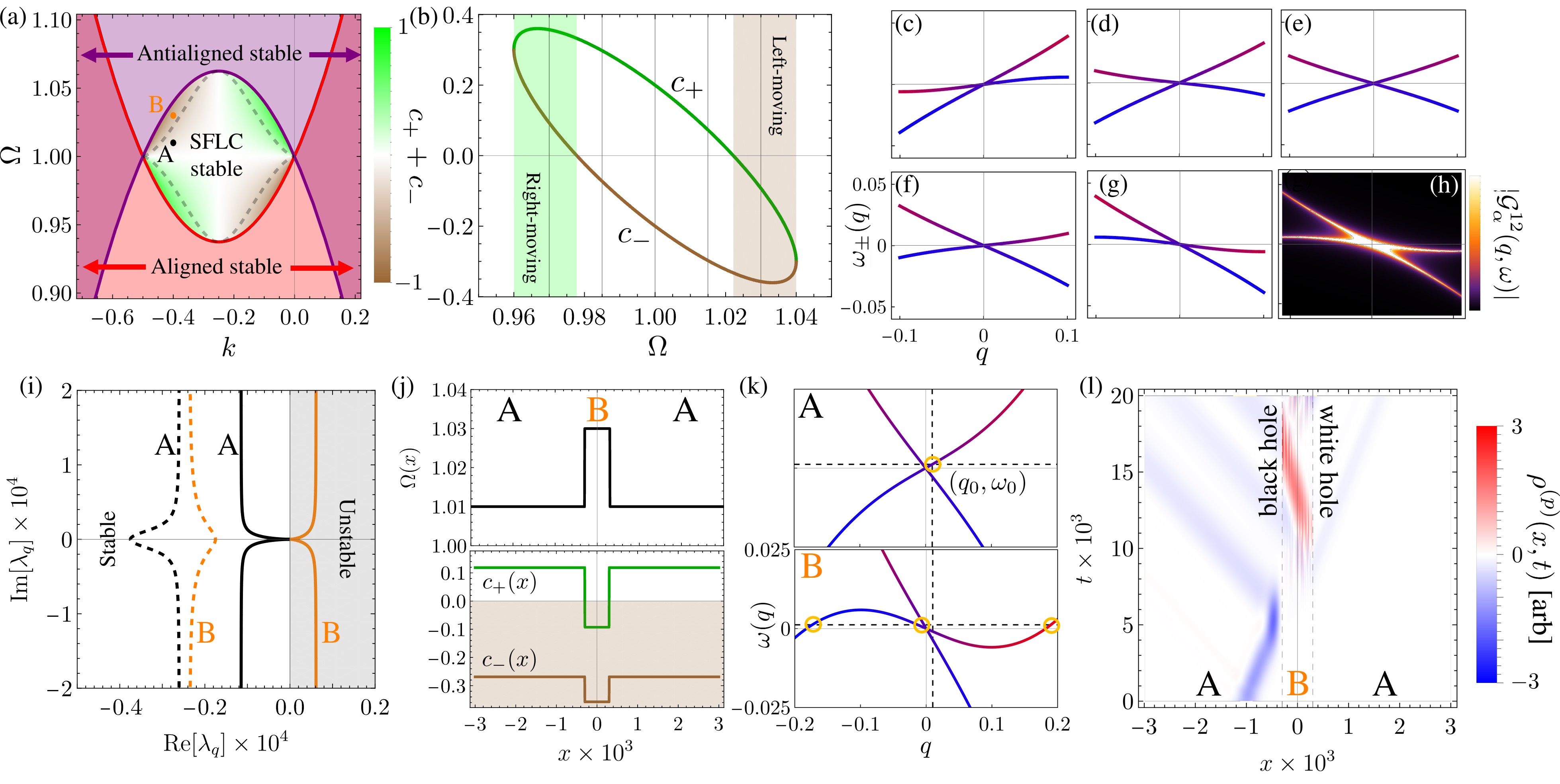}
    \caption{
\textbf{(a)} Wave-propagation asymmetry $c_{+}+c_{-}$ plotted throughout the SFLC stability phase diagram. The central region bounded by gray dashed lines indicates the set of drive strengths $J_{s}=\alpha\Omega$ and pitches $k$ where an SFLC is stably supported. Regions bounded by the purple/red lines and the dashed boundaries correspond to parameter sets where only the Goldstone mode is unstable, or equivalently, long-wavelength spin waves on top of the SFLC propagate unidirectionally. If $(k,J_{s}/\alpha)$ is in the red (purple) region, a spin wave of momentum $k$ will relax toward the aligned (antialigned) state $\mathbf{m}=+\hat{\mathbf{z}}$ ($-\hat{\mathbf{z}}$); in the overlapping region, the final steady state depends on the initial canting angle. The black (A) and orange (B) markers indicate the parameter choices used in panels (i–l). 
\textbf{(b)} Sound velocities $c_{\pm}$ defined below Eq.\,\eqref{eq: asymdisp} versus SFLC frequency $\Omega$; green (brown) curves correspond to $c_{+}$ ($c_{-}$), with shaded regions highlighting right- (left-)propagating unidirectional regimes. 
\textbf{(c–g)} Magnon dispersion atop the SFLC of pitch $k=-0.4$ for $\Omega=0.97,\,0.985,\,1,\,1.015,\,1.03$ (increasing left to right). 
\textbf{(h)} Anomalous correlator strength defined below Eq.\,\eqref{eq: GF} for the $\Omega$ of (g). 
\textbf{(i)} Complex eigenvalue bands $\lambda^{(1)}_{q}$ (solid) and $\lambda^{(2)}_{q}$ (dashed) of the dynamical matrix $\mathbf{M}_{q}$ defined above Eq.\,\eqref{eq: rqsq}, plotted for $q\in[-0.1,0.1]$. The color of each band corresponds to the SFLC parameters A and B in (a). 
\textbf{(j)} Spatial profiles of the local SFLC frequency and sound velocities in the scattering scenario considered, using the same color code as in (b). 
\textbf{(k)} Dispersion within regions A (top) and B (bottom) of the scattering geometry of (j). The color coding matches that in (c–g). The vertical (horizontal) dashed line indicates the momentum $q_{0}$ (frequency $\omega_0=\omega_p(q_{0})$) used for the initial condition Eq.\,\eqref{eq: IC}. The yellow circle in A encloses $(q_{0},\omega_{0})$, while the three yellow circles in B identify the intersections of $\omega_0$ with the dispersion curves in region B, corresponding to the frequency-allowed scattering channels. 
\textbf{(l)} Wavepacket scattering visualized using the particle pseudocharge density defined in Eq.\,\eqref{eq: rhop}. For (a–l), $d=0.25$ and, in (i,l), $\alpha=10^{-3}$. In (k–l), horizons are located at $\pm x_{0}=\pm 200$. The initial condition used in (l) is defined in and around Eq.\,\eqref{eq: IC} with $q_{0}=0.01$ with $\chi$ being a Gaussian centered at $x_{0}=-1000$ and characteristic width $200$.
}
\label{fig: BigFig}
\end{figure*}

Owing to the BdG structure of Eq.~\eqref{eq: dissBdG}, the normal modes (eigenvectors) of $\mathbf{G}_q$ naturally split into particle-like and hole-like sectors. This separation becomes precise upon quantization, with 
$\psi_q \mapsto \sqrt{2s}\,\hat{a}_q$ (``carrying $\hbar$") and $\psi_{-q}^* \mapsto \sqrt{2s}\,\hat{a}^\dag_{-q}$ (``carrying $-\hbar$")
in terms of magnon annihilation and creation operators. These sectors are distinguished by the \textit{Krein signature} $\kappa$ of the corresponding eigenvector, defined as the sign of the indefinite inner product 
$\vec{v}^\dag \bm{\sigma}_3 \vec{v}$ \cite{ColpaQBH_1978,SchulzBaldes_2018,Decon_2020,Antimagnonics_Rembert2024}. In our convention, $\kappa = -1$ $(+1)$ corresponds to the particle (hole) band \footnote{Note that, unlike for fermions, there are no true bosonic holes. A mapping $a\mapsto a^\dag$ reverses the canonical commutation relations, and thus is not unitarily implementable. Nonetheless, the particle-hole language is a convenient shorthand for our BdG description.}.  The particle-band dispersion takes the form
\begin{align}
\omega_p(q) &= -2 m_\text{LC}^z v_k q + |q|\sqrt{q^2 - 2\Delta_k} 
\approx 
\begin{cases} 
c_- q, & q<0,\\[1ex]
c_+ q, & q>0,
\end{cases}\label{eq: asymdisp}
\end{align}
where $c_\pm = -2 m_\text{LC}^z v_k \pm \sqrt{-2\Delta_k}$ is the sound velocity of the right- (left-) moving branch with respect to $q=0$. The hole band is related by
$\omega_h(q) = -\omega_p(-q)$ via the particle-hole constraint. Proceeding under the assumption $\Delta_k<0$, or equivalently $v_k^2<d^2$, we thus uncover generically asymmetric, nonreciprocal dispersion $(c_+\neq - c_-)$  for excitations of negative and positive momenta. In the lab frame, this corresponds to excitations of momentum $q<-k$ and $q>-k$, respectively. In Fig.\,\ref{fig: BigFig}(a) we plot an asymmetry measure $c_++c_-$ throughout the region of SFLC parameter space where $\Delta_k<0$. This measure vanishes (indicating symmetric dispersion) only when $\Omega = 1$, wherein the SFLC is resonant with the applied field, or $k=-d$ corresponding to the zero velocity SFLC background. The sound velocities $c_\pm$ are plotted in Fig.\,\ref{fig: BigFig}(b) while varying the SFLC frequency at a fixed pitch. In Figs.\,\ref{fig: BigFig}(c-g), the long-wavelength dispersion is plotted for various choices of $\Omega$ with clear asymmetry visible in all cases except for (e) where $\Omega = 1$. 

Remarkably, there even exist regimes where the long-wavelength dynamics are completely unidirectional: $c_+c_->0$, shown as green and brown regions of Fig.\,\ref{fig: BigFig}(b). In these regimes, both the particle and hole bands become negative on one side of $q=0$. While in conventional bosonic systems this would indicate a thermodynamic (or Landau) instability \cite{SpinBEC_Ueda2012,Decon_2020}, it does not in this model. That is because the nonequilibrium SFLC background sets an effective ``Fermi level" at $\Omega$ in the lab frame: negative energy excitations represent lab-frame excitations of positive frequencies just below $\Omega$.

Unlike the non-Hermitian skin effect,  which relies on asymmetric hopping~\cite{HatanoNelson_Hatano1996,NHSE_Yao2019,NHSE_Slager2020} (typically induced by a precise combination of coherent and incoherent interaction \cite{Nonrec_Anja2015,DissSWDiode_Zou2024,XinPaper_2025}), the nonreciprocity discussed here arises even though  $\mathbf{G}_q$ generates coherent dynamics with strictly reciprocal couplings. Its origin is the directed motion of the SFLC background, which breaks Galilean invariance: i.e., transforming to the frame comoving with the superflow (effectively setting $v_k=0$ in Eq.\,\eqref{eq: asymdisp}) does not halt magnon propagation. This behavior is characteristic of uniform hydrodynamic states (UHSs) \cite{GalBreak_Ezio2017}, of which the SFLC is a concrete example. Prior discussions confined boost-noninvariant UHSs to undriven, lossless limits, with damping merely curtailing lifetime. Here, inversion-symmetry–breaking DMI, together with a compensating dc drive, stabilizes a manifold of fully stable and locally attractive UHSs — a \textit{bona fide nonequilibrium phase} rather than a fine-tuned transient.

An eigenmode analysis of $\mathbf{G}_q$ reveals further insights. At $q=0$ the particle and hole bands coalesce at an exceptional point (EP) with $\mathbf{G}_{0}$ possessing one eigenvector \(\propto [i,-i]\).  For $\psi_0 = i$, we have $\delta n^x =0$ and $\delta n^y=1$, physically corresponding to a uniform \(y\)-tilt of \(\mathbf{n}\) that advances \(\mathbf{m}\) along the SFLC orbit without changing the pitch. This is the Goldstone mode previously identified with azimuthal phase fluctuations in the lab-frame magnetization $\mathbf{m}$ (see Fig.\,\ref{fig: UnwindFig}(b) and \ref{app: lab}). For $q\neq 0$, $\mathbf{G}_q$ is diagonalizable with two modes whose Nambu (particle–hole) weights continuously approach equal modulus as $q\to 0$, as shown in Figs.\ref{fig: BigFig}(c-g). The  mixing is captured by  the Green's function
\begin{align}\label{eq: GF}
    \mathcal{G}(q,\omega) = (\omega\mathds{1}_2 - \mathbf{G}_q)^{-1},
\end{align}
whose off-diagonal component $\mathcal{G}^{12}(q,\omega)$ measures the anomalous (pair) amplitude. Figure \,\ref{fig: BigFig}(h) shows that the anomalous amplitude peaks at the Goldstone point $(q,\omega)=(0,0)$: 
the SFLC background acts as a parametric pump that mixes particles and holes, generating long-wavelength correlated $(q,-q)$ pairs with, in the quantum regime, two-mode-squeezed statistics. We stress that the classical analysis establishes the parametric pair-generation mechanism itself. Quantum squeezing and the associated entanglement features it engenders, additionally require quantizing the modes and a low thermal occupation.

We now assess the stability of the  eigenmodes by analyzing the dynamical matrix  $\mathbf{M}_q = -i(\mathds{1}_2-i\alpha\bm{\sigma}_3)^{-1}\mathbf{G}_q$, defined implicitly by rearranging Eq.\,\eqref{eq: dissBdG} into the form $\partial_t \Psi_q = \mathbf{M}_q\Psi_q$. A mode fluctuation is dynamically stable when $\mathrm{Re}\,\lambda_q<0$ for its eigenvalue $\lambda_q$. 
At $q=0$ the two eigenvalues satisfy
\begin{equation}
\mathrm{Re}\,\lambda^{(1)}_0=0, 
\qquad
\mathrm{Re}\,\lambda^{(2)}_0=\frac{2\alpha\,\Delta_k}{1+\alpha^2}\,.
\end{equation}
The first, infinite-lifetime, mode is the Goldstone mode which survives in the presence of dissipation. The second mode is stable only when $\Delta_k<0$ - the same condition already noted for dynamical stability of the coherent dynamics generated by $\mathbf{G}_q$.
For small but nonzero \(q\), the second mode remains damped for \(\Delta_k<0\), while the Goldstone mode acquires a perturbative decay rate:
\begin{align}\label{eq: rqsq}
    \text{Re}\,\lambda_q^{(1)} &\approx -\frac{2(m_\text{LC}^z v_k)^2 + \Delta_k}{\alpha \Delta_k}q^2 = \frac{4c_+c_-}{\alpha(c_+-c_-)^2}q^2\,.
\end{align}
Local stability around $q=0$ requires the coefficient of $q^2$ in Eq.\,\eqref{eq: rqsq} to be negative, which translates into the condition $c_+c_-<0$. In this regime, a local phase twist can decompose into left- and right-moving packets that spread and decay under Gilbert damping. On the other hand, in the unidirectional regime \(c_{+}c_{-}>0\), phase twists cannot disperse by counter-propagation; the anomalous coherence pumped by the SFLC then accumulates, and the Goldstone mode rate changes sign, signaling a parametric instability. In the following, we will show that the stable-to-unstable transition at \(c_{+}c_{-}=0\) can be also understood as the kinematic fingerprint of black-hole horizon, with Hawking-like pair production.

\section{Long-wavelength analogue \\ black hole horizons}
To make the analogue-gravity parallel explicit, we recast the long-wavelength dynamics in an acoustic-metric form. In the long-wavelength limit, the Goldstone mode $\delta n^y$ (or equivalently in the lab-frame $\delta\phi$ - see Appendix \ref{app: lab}) obeys the hyperbolic wave equation  
\begin{align}\label{eq: waveeq}
    (\partial_t - c_-\partial_x)(\partial_t-c_+\partial_x)\delta n^y = 0\,,
\end{align}
whose characteristics correspond to the null trajectories of the effective spacetime line element
\begin{align}\label{eq: lineelement}
    ds^2 = c_+c_- dt^2 -(c_++c_-)dx dt +dx^2.
\end{align}
Equation \eqref{eq: lineelement} reveals two main  intertwined kinematic knobs: (i) the product \(c_{+}c_{-}\) fixes the light-cone opening and flips sign at a horizon (\(c_{\pm}=0\)), marking the transition from bi- to unidirectional propagation; (ii) 
the shift term \(c_{+}+c_{-}\propto v_{k}\) 
   encodes flow-induced frame dragging by the chiral superfluid background.

These observations motivate an analogue–gravity construction in which Hawking-like emission in a tabletop magnet is realized by spatially modulating the spin–transfer torque. To make this concrete, we consider the scattering geometry depicted in Fig.\,\ref{fig: BigFig}(j). A smooth profile \(J_s(x)\) -- and thus \(\Omega(x)=J_s(x)/\alpha\) -- imprints position dependence on the sound speeds \(c_\pm(x)\) and selects locations \(x=\pm x_0\) where the right-moving branch closes, \(c_+(\pm x_0)=0\), establishing sonic horizons that separate propagation regimes. Outside \(|x|>x_0\) (region A), \(J_s(x)\approx J_s^A\) is nearly constant and tuned to the bidirectional phase with \(c_+(x)c_-(x)<0\); inside \(|x|<x_0\) (region B), \(J_s(x)\approx J_s^B\) is nearly constant and tuned to the unidirectional phase with \(c_+(x)c_-(x)>0\). At the interfaces \(x=\pm x_0\), we take \(J_s(x)\) to smoothly interpolate between $J_s^A$ and $J_s^B$ over a characteristic length scale \(l\).

Suppose Eq.\,\eqref{eq: LLG} admits a solution that supports, within region A (B), an SFLC with pitch $k_A$ $(k_B)$ and frequency $\Omega_A=J_s^A/\alpha$ ($\Omega_B=J_s^B/\alpha$) with dynamical domain walls in the interfacial regions. In such a scenario, magnon transport can be analyzed by suitably generalizing Eq.\,\eqref{eq: dissBdG}. In real space, the equation of motion reads 
\begin{align}\label{eq: dissBdGx}
i(\mathds{1}_2 - i\alpha \bm{\sigma}_3)\,\partial_t \Psi(x,t) = \mathbf{G}(x)\Psi(x,t),
\end{align}
where $\mathbf{G}(x)$ is obtained from $\mathbf{G}_q$ in Eq.\,\eqref{eq: Gq} by replacing $q \mapsto -i\partial_x$ and $J_s \mapsto J_s(x)$. Explicitly, with $k_{A,B}=k$,
\begin{align}
    \mathbf{G}(x) = \Delta_k(x)(\bm{\sigma}_3+i\bm{\sigma}_2) +2im_\text{LC}^z(x) v_k \mathds{1}_2\partial_x + \bm{\sigma}_3 \partial_x^2.
    \label{eq: G(x)}
\end{align}
It is important to note that the above equation is exact only when the SFLC precesses locally at \(\Omega(x)=J_s(x)/\alpha\), in which case it must be interpreted pointwise in a frame rotating at \(\Omega(x)\). Rapid spatial variations of \(J_s(x)\) generate additional, non-negligible terms from the position-dependent rotation \(\mathbf{R}_z[-\Omega(x)t]\) in the unwinding map Eq.\,\eqref{eq: unwind}, i.e., \(\propto \partial_x\Omega\), which are neglected here. Thus, our analysis is valid everywhere except within a narrow interfacial region. In this spatially adiabatic regime, a component of $\Psi(x,t)\propto \chi_\omega(x)e^{-i\omega t}$ localized near $x$ corresponds in the lab frame to an excitation at frequency $\Omega(x)+\omega$ -that is, $\omega$ always represents a local detuning.

To understand scattering dynamics, within each A/B region, we can define the momentum-space BdG Hamiltonian $\mathbf{G}_q^{A/B} = \mathbf{G}_q\big|_{J_s=J_s^{A/B}}$, whose spectra are shown in Fig.\,\ref{fig: BigFig}(k). Due to time-translation invariance of Eq.\,\eqref{eq: dissBdGx} (implying frequency conservation), an incident right-moving particle mode  with momentum $q_0>0$ and frequency $\omega_0=\omega^{(p),A}(q_0)>0$ (dashed lines in dispersion A) can only scatter into (i) long-wavelength left-moving particle and hole states (central yellow circle in dispersion B) or (ii) short-wavelength right-moving particle and hole states (left- and rightmost yellow circles in dispersion B).

To test the acoustic-metric picture of Eqs.\,\eqref{eq: waveeq}-\eqref{eq: lineelement},  we launch a right-moving, long-wavelength wave packet from the left asymptotic region ($x \ll -x_0$) with central momentum $q_0 \ll 1$. Specifically, we write $\Psi(x,0) = \Psi^{(p)}(x) + \Psi^{(h)}(x)$ where the particle component is given by 
\begin{align}\label{eq: IC}
    \Psi^{(p)}(x) &= \chi(x) e^{iq_0 x} \mathbf{u}_{q_0}^A\,,
\end{align}
with $\chi(x)$ a Gaussian localized at $x=x_0\ll -x_0$ and  $\mathbf{u}_{q_0}^A$ is the eigenvector of $\mathbf{G}_q^A$ at $q=q_0$ with frequency $\omega_p(q_0)>0$ satisfying the pseudo-Hermitian normalization $\mathbf{u}_{q_0}^{A}{}^\dag \bm{\sigma}_3 \mathbf{u}_{q_0}^A = -1$. Reality of $\delta n^x$ and $\delta n^y$ fixes the hole partner as $\Psi^{(h)} = \bm{\sigma}_1 \Psi^{(p)}(x)^*$. 
Linearity allows us to evolve the particle and hole components independently and the relationship $\mathbf{G}^*(x) = -\bm{\sigma}_1\mathbf{G}(x)\bm{\sigma}_1$ guarantees $\Psi^{(h)}(x,t) = \bm{\sigma}_1 \Psi^{(p)}(x,t)^*$ for all $t\geq 0$. 

As a quantitative witness to particle-hole dynamics, we define the particle pseudocharge density
\begin{align}\label{eq: rhop}
    \rho^{(p)}(x,t) = \Psi^{(p)}{}^\dag(x,t)\bm{\sigma}_3\Psi^{(p)}(x,t).
\end{align}
For our initial state, $\rho^{(p)}(x,0) = -\chi^2(x)$. In general, $\rho^{(p)}(x,t)<0 (>0)$ indicates a concentration of particles (holes) at $(x,t)$ and it flips sign only when particle–hole correlations are generated (see \ref{app: magcont}). 

Figure \,\ref{fig: BigFig}(l) shows the evolution of $\rho^{(p)}(x,t)$ corresponding to the scattering of the Gaussian wavepacket with particle component Eq.\,\eqref{eq: IC}  through the geometry depicted in Fig. \ref{fig: BigFig}(j). Initially, the long-wavelength particle density (blue color) propagates toward the black-hole horizon with group velocity  $d\omega_p/dq\big|_{q=q_0}>0$. At the horizon, long-wavelength particles  are unable to tunnel and instead weakly scatter into right-moving higher momentum particle and hole channels within region B. Upon reaching the white-hole horizon, these modes split: a portion reflects as long-wavelength holes back into 
region B, while the remainder transmits as long-wavelength particles into region A. The black-hole--white-hole pair thus forms a black-hole-laser cavity that repeatedly squeezes the $(q,-q)$ pair, 
generating quasi-periodic, Hawking-like particle bursts into the left-half of region A accompanied by sign reversals of $\rho^{(p)}$ that mark particle $\leftrightarrow$ hole conversion. In parallel, a faint, phase-coherent thread of low-$q$ particles traverses B and emerges into the right half of A -- a BdG Klein-tunneling analogue.

Several recent studies have proposed magnetic analogues of curved spacetime and relativistic scattering, including magnonic black holes and Klein tunneling in spin torque-driven or chiral magnetic media. Early proposals demonstrated that a uniform current can Doppler-shift magnon modes to create spin flow horizons when the drift velocity equals the spin wave speed \cite{Horizon_Rousseaux2011,MagBH_Rembert2017}, while subsequent works explored interface-based antimagnonic amplification \cite{MagKlein_Rembert2022}, lattice analogues of Dirac fields on curved backgrounds \cite{ChiralBH_Forbes2023}, cavity-like black-hole lasing driven by strong spin torques in patterned waveguides \cite{MagAmplif_Nakayama2024}, and even spin superfluid analogues in superfluid \textsuperscript{3}He-B \cite{MagSFHe_Skyba2019}. These approaches rely on externally imposed currents or engineered inhomogeneities to kinematically generate horizon conditions within near-equilibrium magnets. By contrast, our mechanism embeds the analogue-gravity physics inside a self-organized nonequilibrium phase: the chiral SFLC arises dynamically from the interplay between drive and damping, producing an intrinsically nonreciprocal background. A smooth spatial modulation of the spin torque then maps onto an acoustic metric and creates sonic horizons—without externally provided flows or fragile gain–loss fine-tuning. This self-oscillatory route unifies nonreciprocity, analogue gravity, and driven-dissipative phases within a single, experimentally transparent magnetic platform.

\section{Discussion}

We have shown that a conventional multilayer ferromagnet, operated under a steady spin torque drive, supports a chiral spin superfluid limit cycle that spontaneously breaks continuous time and space translations and endows long-wavelength magnons with an intrinsically nonreciprocal dispersion. In the comoving, corotating frame, the linearized dynamics are pseudo-Hermitian, and the long-wavelength Goldstone mode obeys a
scalar wave equation with a (1+1)D acoustic metric. Spatially patterned drives thus act as a kinematic control knob for sonic horizons and Bogoliubov mixing, enabling Hawking-like particle–hole pair production. 

Mechanistically, the nonreciprocity is flow-borne: it originates from the directed SFLC background and the concomitant breaking of Galilean invariance, not from asymmetric couplings or fragile gain--loss balance. This separates our diode effect from phenomena such as the non-Hermitian skin effect, where asymmetric hopping is essential. In our case, transforming to the superflow frame does not freeze propagation; the two sound speeds remain distinct, and their product and sum control the “light-cone” opening and effective frame-dragging of the acoustic geometry. The emergence of an exceptional point and a 
Goldstone mode provide a unified lens on both the time-crystalline-like order and the particle–hole structure of fluctuations.

These kinematic handles translate into practical diagnostics. 
The flow-set asymmetry of the long-wavelength branches implies direction-selective transmission that can be read out with standard spintronics probes.
Entering the unidirectional regime \(c_{+}c_{-}>0\) pins one chirality and produces threshold-like amplification of correlated \((q,-q)\) pairs -- signals that should appear as drive-tunable sidebands, excess low-frequency noise, and phase-sensitive two-mode correlations. 
In horizon geometries, \(\mathrm{SU}(1,1)\) Bogoliubov mixing converts incident long-wavelength particles into high-\(q\) modes and back into long-wavelength particle--hole partners, yielding quasi-periodic emission bursts and sign flips of a particle--hole witness (pseudocharge) across the interface. 
Each of these observables follows directly from the pseudo-Hermitian BdG structure and the acoustic-metric hydrodynamics derived here.

Conceptually, this platform broadens the landscape of nonreciprocal phases and nonequilibrium criticality. Rather than approaching nonreciprocal fixed points via static order and critical exceptional points (CEPs), we reach a chiral, time-crystalline state by a dynamical bifurcation from a steady driven phase, after which spectral topology (i.e., exceptional degeneracies) and hydrodynamics together govern transport.
Experimentally, the ingredients are standard -- Gilbert damping, interfacial exchange and DMI, and dc spin current injection -- and the key control parameters are transparent. This simplicity suggests immediate routes in sputtered multilayers where interlayer exchange and DMI are already engineered and characterized~\cite{DMIAFM_Fern2019,LongRangeChiralAFM_Han2019,ChiralCoup_Avci2021} and the phenomena we predict can be resolved via Brillouin light scattering, broadband ferromagnetic resonance, magnetization-noise and  propagating spin wave spectroscopy~\cite{flebus20242024}, and--on the microsecond timescales relevant here--through all-electrical detection schemes based on anisotropic magnetoresistance (AMR), where the induced rotations of the magnetizations generate measurable voltage signals \cite{AMRExp_Schumacher2007}. Time-resolved MOKE pump–probe measurements provide an additional complementary pathway \cite{Moke_Kirilyuk2010}.

Several effects beyond our minimal model, e.g., interlayer dipolar interactions and magnon–magnon interactions beyond the quadratic fluctuation analysis, merit future study. We expect the former to largely renormalize the parameters of Eq.\,\eqref{eq: Fdiml} while preserving the U(1) superfluid phase, and the latter to dress the fluctuation spectrum and saturate the parametric amplification without altering the flow-borne origin of the nonreciprocity established here.

Taken together, these results position magnetic heterostructures as a quantitative testbed for non-equilibrium universality where nonlinearity and non-Hermiticity are independently tunable and experimentally accessible.
Furthermore, the emergence of nonreciprocity and a phase Goldstone mode links our system to recently uncovered nonreciprocal transitions at non-Hermitian quantum criticality \cite{CritEP_Hanai2020,NonRec_Vitelli2021,CEP_Diehl2024,NRIsing_Vitelli2025}, yet the mechanism is distinct: the transition into the SFLC state is not facilitated through a CEP, nor is it from a static already symmetry-broken state. These parallels and differences call for a systematic comparison of the two routes to criticality.

\acknowledgements

The authors would like to thank Lorenza Viola, Michiel Burgelman, and Joeseph Sklenar for fruitful discussions. 
B.F. was supported by the National Science Foundation under Grant No. NSF DMR-2144086.

\appendix

\section{SFLC dynamics in a discrete heterostructure}\label{app: num}
In this Appendix, we present the discrete analogue of our model in Eqs.\,\eqref{eq: LLG} and \eqref{eq: Fdiml} and demonstrate the existence of an SFLC attractor under realistic device conditions. Specifically, consider the explicit magnetic free energy of the (discrete) heterostructure in Fig.\,\ref{fig: SFLC_cartoon}:
\begin{align}
    F_\text{d} = \sum_{j=1}^N\left( -J \mathbf{m}_j\cdot \mathbf{m}_{j+1} + D\hat{\mathbf{z}}\cdot(\mathbf{m}_j\times \mathbf{m}_{j+1}) - M_s B_0 m^z_j\right).
\end{align}
where $J$ and $D$ are the layer-mediated symmetric and antisymmetric strengths, respectively, and $B_0$ is the applied field along $\hat{\mathbf{z}}$. In the above, periodic boundary conditions (PBCs) and open boundary conditions (OBCs) may be imposed via $\mathbf{m}_{N+1} = \mathbf{m}_1$ and $\mathbf{m}_{N+1} = 0$, respectively. Upon introducing a lattice constant $a$, taking the continuum limit $\mathbf{m}_j(t) = \mathbf{m}(x=ja,t)$, we arrive at the model of the main text with the identifications
\begin{align}
    A = Ja,\quad B = \frac{B_0}a,\quad d = \frac{D}{\sqrt{J M_sB_0}}.
\end{align}
Under PBCs, one may show that the LLG equation supports a discrete LC solution which, in terms of the azimuthal angles $\phi_j$ and $z$ spin components, is
\begin{subequations}
\begin{align}
    \phi_j(t) &= k_m j+\Omega t,
    \\
    m^z_j &= \frac{1-\Omega/\omega_0}{2\left[ \bar{J}(\cos(k_m)-1) - \bar{D}\sin(k_m)\right]} \label{eq: mzdisc},
\end{align}
\end{subequations}
where $k_m = 2\pi n/N$ with $n\in\mathbb{Z}$, $\Omega = \omega_0 J_s/\alpha$, $\omega_0=\gamma B_0$, and $\bar{J} = J/(M_sB_0)$, $\bar{D}=D/(M_s B_0)$ are the dimensionless exchange strengths. Stability of an SFLC with pitch $k_m$ requires that
\begin{align}
    \left|1-\frac{J_s}{\alpha\omega_0}\right| < 2\left[\bar{J}(\cos(k_m)-1)-\bar{D}\sin(k_m)\right],
\end{align}
constraining the applied drive within a set $k_m$-dependent range around the characteristic decay rate $\alpha \omega_0$.

\begin{figure*}
    \centering
    \includegraphics[width=1.0\linewidth]{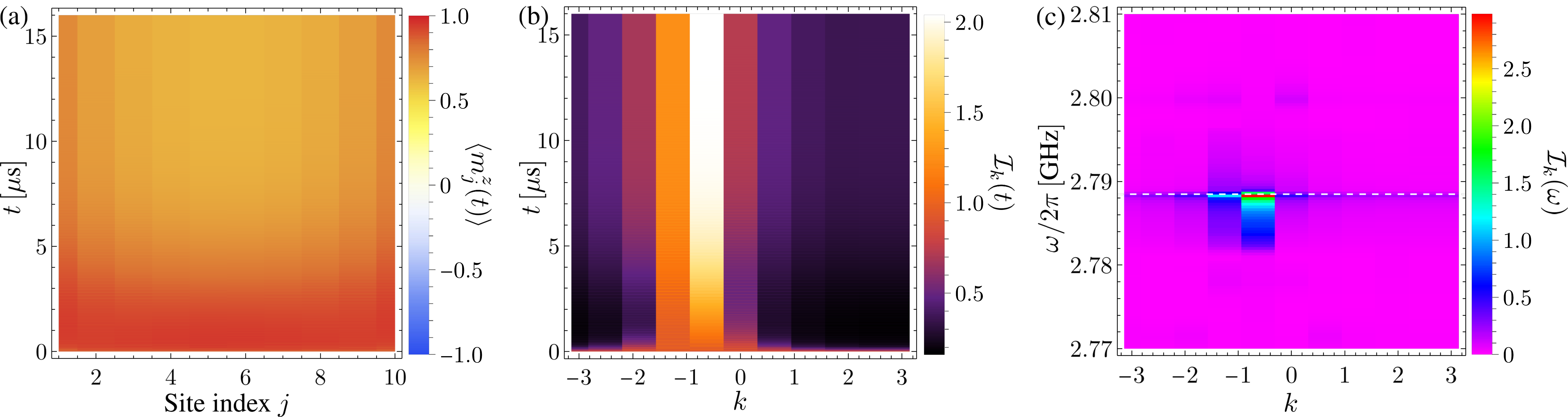}
    \caption{Demonstration of the SFLC state in a realistic heterostructure. \textbf{(a)} The ensemble average of the $z$-component of the macrospins $\mathbf{m}_j(t)$. \textbf{(b)} The real-time momentum resolved transverse spin density Eq.\,\eqref{eq: ktdens}. \textbf{(c)} The frequency- and momentum-resolved transverse spin density Eq.\,\eqref{eq: komdens}. The white dashed line is the predicted SFLC frequency $\Omega_\text{LC} = J_s/\alpha$. In all cases, $N=10$ layers are simulated under OBCs under an applied field $B_0 = 100\,$mT, with layer-mediated symmetric and antisymmetric exchange fields $J/M_s = D/M_s = 1$ mT. The Gilbert damping is $\alpha = 0.01$ and the applied drive is $J_s = 0.995\alpha\omega_0$.}
    \label{fig: appfig}
\end{figure*}

Real heterostructures feature translational symmetry-breaking OBCs, and thus the above analytical solution is no longer valid. However, we find, via numerical simulation, that indeed, SFLC-like spatiotemporal order emerges in steady state. In Fig.\,\ref{fig: appfig}, we plot example trajectories and ensemble averaged frequency- and momentum-resolved quantities for representative initial conditions and realistic device parameters \cite{DMIObs_Di2015,LongRangeChiralAFM_Han2019,SOTSwitchingSAF_Wang2021,LargeDMI_Arregi2023,UnidirectMagnon_Rembert2023,InterlayerDMI_Zhao2025}. Specifically, we consider an $N=10$ layer heterostructure subjected to an applied field of strength $B_0 = 100\,\text{mT}$. The symmetric and antisymmetric exchange fields are set to $J/M_s=D/M_s = 1\,\text{mT}$, corresponding to $d = \bar{D}/(\bar{J})^{1/2} = 0.1$. The drive strength is set by $J_s = 0.995\alpha \omega_0$ with $\alpha = 0.01$ the Gilbert damping. The LLG equation is solved numerically for 100 initial conditions generated by sourcing $\mathbf{m}_j(0)$ from a uniform distribution on the northern cap of the sphere defined by polar angles $\theta_j\in[0,\pi/4]$. 

In \ref{fig: appfig}(a) we show the dynamics of the ensemble averaged $z$-components $\braket{m_j^z(t)}$. We observe convergence to $\sim 0.25$ in the bulk with some minor variations near the boundaries. In \ref{fig: appfig}(b), we show the ensemble averaged, momentum-resolved, real-time dynamics. Specifically, we plot the quantity
\begin{subequations}
\begin{align}\label{eq: ktdens}
    \mathcal{I}_{k}(t) &= \log\left[ 1 + \braket{|m^+_{k}(t)}|\right],\\ 
    m_k^+(t) &= \sum_{j=1}^Nm_j^+(t) e^{-ikj},
\end{align}
\end{subequations}
as a function of $k$ and $t$. Concentration of $\mathcal{I}_k(t)$ at a given $k=2m\pi/N$ represents the presence of spatial order of characteristic wavelength $\lambda = N/m$ at that time $t$. We observe a clear emergence of spatial order at nonvanishing wavectors $k$ centered around $k_{-1} = -2\pi/N$ in steady state. Inserting this value for $k_m$) into the PBC-derived Eq.\,\eqref{eq: mzdisc}, we obtain $m_j^z \sim 0.630$, which differs from the observed (in Fig.\,\ref{fig: appfig}(a)) steady state bulk value $m_{j=5}^z \sim 0.624$ by a percent error $\sim 0.952\%$. Furthermore, mapping back to the continuum theory, we find that, within this parameter regime and for an SFLC with pitch $k_{-1}$, the degree of nonreciprocity is given by $|c_-/c_+|\sim 4.708$.

Finally, \ref{fig: appfig}(c) shows the quantity
\begin{subequations}
\begin{align}\label{eq: komdens}
    \mathcal{I}_{k}(\omega) &= \log\left[ 1 + \braket{|m^+_{k}(\omega)}|\right],\\
    m_k^+(\omega) &= \int_{t_i}^{t_f} \frac{dt}{2\pi}m_k^+(t) e^{i\omega t},
\end{align}
\end{subequations}
which encodes the ensemble average of both the frequency and momentum content of a given trajectory within the time range $[t_i,t_f]$. Here we choose $t_i = 5\,\mu\text{s}$  after which the LC has set in, and $t_f = 15.1\,\mu$s is the endtime of the simulation. In addition to the emergence of spatial order at $k\sim k_{-1}$, temporal order manifests in large Fourier amplitudes at a nonzero frequency $f\sim 2.78$ GHz which differs from the predicted SFLC frequency $f_\text{LC}$ (white dashed line in Fig.\,\ref{fig: appfig}(c)) by a percent error of $\sim 5\times 10^{-3}\,\%$.

\section{Equilibrium analysis}\label{app: eqanalysis}
In this Appendix we summarize the equilibrium solutions of the functional $\tilde F$ defined in Eq.\,\eqref{eq: Fdiml}.
Besides the fully polarized state, $\mathbf{m}=\hat{\mathbf{z}}$ (with dimensionless free-energy density $-1$), there exists a one-parameter family of spiral configurations with $\phi(x)=kx$ and uniform longitudinal component $m^{z}=-1/f_k$ (admissible for $|f_k|>1$) that satisfy the Euler–Lagrange condition $\delta \tilde F/\delta \mathbf{m}=0$. All such states have energy density given by $(f_k+f_k^{-1})/2$. The condition $|f_k|>1$ constrains $k$ to satisfy either (i) $|k+d|<\sqrt{d^2-1}$ or (ii) $|k+d|> \sqrt{d^2-1}$.
The former class of states, which exist only for $|d|>1$, have energy density strictly less than that of the polarized state. These are the LESSs discussed in the main text. Within this class, the LESS with $k=-d$ has the lowest energy. In contrast, states in the latter class  always exist and have energy density higher than the polarized state. These are the HESSs discussed in the main text.

To determine the energetic stability of these states, we expand the free energy\,\eqref{eq: Fdiml} in fluctuations of the azimuthal angle $\phi$ and the polar angle $\theta = \arccos m^z$. Expanding around the polarized state  $\theta = 0$ leads to fluctuations
\begin{align}
    \delta \tilde{F} = \frac{1}{2}\int \left[ (\partial_x\theta)^2 + (f_\phi+1)\theta^2\right]\,dx\,,
\end{align}
where $f_\phi = (\partial_x\phi)(\partial_x\phi+2d)$ is the free energy density carried by an arbitrary azimuthal profile $\phi(x)$ (which is ill-defined in the polarized equilibrium state). Evidently, the mass gap closes and the polarized state destabilizes when the spiral pitch $\partial_x\phi = k$ satisfies $f_k<-1$, corresponding to LESSs. 

The free energy of fluctuations about an arbitrary spiral state, up to constant shifts, reads
\begin{align}
    \delta \tilde{F} = \frac{1}{2}\int \Big[&(\partial_x\tilde{\theta})^2 -f_kn_k \tilde{\theta}^2 + n_k (\partial_x\tilde{\phi}+v_k)^2 
    \\
    &- \frac{4v_k\sqrt{n_k}}{f_k}\tilde{\theta}\partial_x \tilde{\phi}\Big] dx \,, \label{eq: SSfluc}
\end{align}
where $\tilde{\phi} = \phi - kx$ and $\tilde{\theta} = \theta - \arccos(-1/f_k)$ are the fluctuations in $\phi$ and $\theta$ respectively, $n_k = 1-1/f_k^2>0$ is the superfluid density, and $v_k = k+d$ is the superfluid velocity (such that $j_k=n_k v_k$ is the superfluid current). From left to right, the terms in Eq.\,\eqref{eq: SSfluc} correspond to the polar angle stiffness, the polar angle mass term, the azimuthal stiffness in the presence of the effective superfluid gauge field, and a coupling term between the polar field and the azimuthal gradient. As expected, energetic stability requires that $f_k<0$ (thus excluding stability of HESSs) -- but this criterion is only necessary. To obtain a sufficient condition, we move to Fourier space $(\theta(x),\phi(x))\mapsto \Phi_q = (\theta_q,\phi_q)$ and rewrite Eq.\,\eqref{eq: SSfluc} as 
\begin{subequations}
\begin{align}
    \delta \tilde{F} &= \frac{1}{2}\int \Phi_q^\dag \mathbf{F}_q \Phi_q\,dq  +\delta \tilde{F}_\text{bdry},
    \\
    \mathbf{F}_q &= \begin{bmatrix}
n_k q^2 & 2i v_k \sqrt{n_k}q/f_k
\\
- 2i v_k \sqrt{n_k}q/f_k & q^2-f_kn_k\,,
    \end{bmatrix}
\end{align}
\end{subequations}
where $\delta\tilde{F}_\text{bdry}$ is a boundary term due to the gauge field. Energetic stability is equivalent to positive definiteness of the Hermitian matrix $\mathbf{F}_q$, which implies the following necessary and sufficient conditions for energetic stability of a LESS:
\begin{align}
    f_k <0\text{ and } v_k^2<\frac{n_k|f_k|^3}{4} =\frac{n_k|v_k^2-d^2|^3}{4} .
\end{align}
The upper bound on the superflow velocity  is known as the Landau criterion for stability of spin superfluids \cite{SFReview_Sonin}.
 Figure~\ref{fig: EqPD} in the main text shows that some LESSs are energetically unstable. In contrast, the special spiral with $k=-d$ satisfies $v_{k=-d}=0$ and is always energetically stable. Because this configuration also minimizes the energy density within the LESS branch, it is the ground state for $|d|>1$.

\section{Pseudocharge conservation for 
pseudo-Hermitian systems}\label{app: magcont}
In this Appendix, we discuss in detail the definition of the particle pseudocharge density $  \rho^{(p)}(x,t)$~\,\eqref{eq: rhop} used to quantify particle–hole dynamics in the scattering scenario discussed in the main text. For a homogeneous antidamping torque profile $J_s$, the real-space BdG Hamiltonian $\mathbf{G}(x)$~\eqref{eq: G(x)} is pseudo-Hermitian with respect to the indefinite inner product $\braket{\Psi,\Phi}=\int \Psi^\dag(x)\bm{\sigma}_3\Phi(x)dx$. Correspondingly, the local pseudocharge density $\rho(x,t) = \Psi^\dag(x,t)\bm{\sigma}_3\Psi(x,t)$ satisfies a continuity equation $\partial_t\rho + \partial_x \mathcal{J} = 0$ with
\begin{align}
    \mathcal{J} = -2im^z_\text{LC} v_k\Psi^\dag\bm{\sigma}_3\Psi -i\left[ (\partial_x\Psi^\dag) \Psi - \Psi^\dag (\partial_x\Psi)\right].
\end{align}
When $\mathbf{G}(x)$ depends on position through the spin current profile $J_s(x)$ as in Eq.\,\eqref{eq: G(x)}, the coupling between  position and momentum $(-i\partial_x)$ breaks strict pseudo-Hermiticity. Consequently, the continuity law picks up a source: $\partial_t\rho + \partial_x \mathcal{J} = \mathcal{S}(x)$, with $\mathcal{S}(x) \propto J_s'(x)$, signifying that pseudocharge is created or absorbed where $J_s(x)$ varies, e.g., in our case, at the interfaces. For general states, the sign of $\rho$ is tied to particle or hole concentration (with $\rho>0$ ($<0$) indicating particle (hole) concentration), while for physical states obeying the reality constraint $\Psi(x,t) = \bm{\sigma}_1\Psi^*(x,t)$,  the density vanishes, i.e., $\rho(x,t)=0$ for all $(x,t)$. The continuity law then reduces to $\partial_x \mathcal{J} = \mathcal{S}(x)$, implying that the integrated source enforces a jump in the pseudocurrent across an interface. 

Since we wish to quantify particle-hole dynamics (which $\rho=0$ trivially cannot bear witness to), we instead use the particle pseudocharge density defined in Eq.\,\eqref{eq: rhop} which computes $\rho$ for the particle component of a physical state $\Psi$ only.

\section{Mapping lab-frame angle fluctuations to SFLC frame transverse fluctuations}\label{app: lab}

In this Appendix, we relate fluctuations in the lab-frame azimuthal $\phi$ and polar $\theta$ angles to transverse fluctuations of the SFLC-frame magnetization $\mathbf{n}$. Recalling the definition of the SFLC density $n_k = 1-m_\text{LC}^z{}^2 = \sin^2\theta_\text{LC}$, and by expanding both sides of $\mathbf{n}(x,t)=\mathbf{R}(x,t)\mathbf{m}(x,t)$ about the SFLC, we find
\begin{align}\label{eq: phasetrans}
    \delta n^x = \delta\theta,\quad \delta n^y = \sqrt{n_k}\delta\phi\,,
\end{align}
where $\delta n^x$ and $\delta n^y$ are the transverse fluctuations in $\mathbf{n}$, $\delta\theta=\theta-\theta_\text{LC}$, and $\delta\phi = \phi-\phi_\text{LC}$. With this, it becomes natural to associate the real and imaginary quadratures associated with $\psi=\delta n^x+i\delta n^y$ with polar and azimuthal fluctuations, respectively. In particular, the Goldstone mode $\delta\phi$ is simply the imaginary quadrature $\text{Im}[\psi]$ scaled by the SFLC density $n_k$.
The dissipative BdG equation~\,\eqref{eq: dissBdG} accordingly maps to
\begin{align}
i(\mathds{1}_2+i\alpha\bm{\sigma}_2)\partial_t\Phi_q = \tilde{\mathbf{G}}_q \Phi_q\,,
\end{align}
where  $\Phi_q = (\delta\theta_q,\sqrt{n_k}\delta \phi_q)$ and $\tilde{\mathbf{G}}_q = -2 m_\text{LC}^z v_k  q\mathds{1}_2 +(q^2-\Delta_k) \bm{\sigma}_2 -  \Delta_k i\bm{\sigma}_1$ is unitarily equivalent to $\tilde{\mathbf{G}}_q$. When $q=0$, for example, $\tilde{\mathbf{G}}_0$ has one eigenvector $[0,1]$ at zero frequency - the Goldstone mode. In the formal limit $\alpha \to 0$, this system of first-order ODEs (in time)  can be mapped onto a single second-order ODE 
\begin{align}
    (\partial_t +i c_+q)(\partial_t+ic_-q)\delta\phi = \mathcal{O}(q^4),
\end{align}
which, in the limit $q\to 0$, and under the identification Eq.\,\eqref{eq: phasetrans}, leads to Eq.\,\eqref{eq: waveeq} in the main text.

\end{document}